\documentclass{article}
\usepackage{graphicx}%
\usepackage{multirow}%
\usepackage{mathrsfs}%
\usepackage[title]{appendix}%
\usepackage{xcolor}%
\usepackage{textcomp}%
\usepackage{manyfoot}%
\usepackage{algorithm}%
\usepackage{algorithmicx}%
\usepackage{algpseudocode}%
\usepackage{listings}%

\usepackage{arxiv}
\usepackage{latexsym}              % symbols
\usepackage{amsmath}               % great math stuff
\usepackage{amssymb}               % great math symbols
\usepackage{amsthm}                % for theorems, 
\usepackage{dsfont}
\usepackage[utf8]{inputenc} % allow utf-8 input
\usepackage[T1]{fontenc}    % use 8-bit T1 fonts
\usepackage{hyperref}       % hyperlinks
\usepackage{url}            % simple URL typesetting
\usepackage{booktabs}       % professional-quality tables
\usepackage{amsfonts}       % blackboard math symbols
\usepackage{nicefrac}       % compact symbols for 1/2, etc.
\usepackage{microtype}      % microtypography

\usepackage[sort,numbers,super]{natbib} %superscript citation for Nature
\usepackage{doi}
\usepackage{color}
\usepackage{tcolorbox}
\usepackage{tikz}                  % to make drawing with TikZ
\usetikzlibrary{arrows,positioning,shapes,bayesnet}
\usepackage{flafter}
\usepackage{pdfpages}
\usepackage{cleveref}
\crefname{assumption}{Assumption}{Assumptions}
\crefname{equation}{Eq.}{Eqs.}
\crefname{figure}{Fig.}{Figs.}
\crefname{table}{Table}{Tables}
\crefname{section}{Sec.}{Secs.}
\crefname{theorem}{Thm.}{Thms.}
\crefname{lemma}{Lemma}{Lemmas}
\crefname{corollary}{Cor.}{Cors.}
\crefname{example}{Example}{Examples}
\crefname{appendix}{Appendix}{Appendixes}
\crefname{remark}{Remark}{Remark}

\newcommand{\paren}[1]{\left( #1 \right)}

\newcommand{\I}{\mathcal{I}}
\newcommand{\J}{\mathcal{J}}

\newcommand\sigmaij{{\sigma_i^j}^2}

\title{Longitudinal prediction of DNA methylation to forecast epigenetic outcomes}

\date{19 December 2023}	% Here you can change the date presented in the paper title
\author{ \href{https://orcid.org/0000-0003-0806-8934}{\includegraphics[scale=0.06]{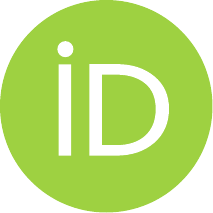}\hspace{1mm}}Arthur Leroy\\ %\thanks{Use footnote for providing further
	Department of Computer Science\\
	The University of Manchester\\
	\texttt{arthur.leroy.pro@gmail.com} \\
	\And
	Ai Ling Teh \\
	Singapore Institute for Clinical Sciences (SICS) \\
        Bioinformatics Institute (BII) \\
        Agency for Science, Technology and Research (A*STAR) \\
	\texttt{ailing\_teh@sics.a-star.edu.sg} \\
	  \AND
	Frank Dondelinger \\
	  Lancaster Medical School \\
	  Lancaster University \\
	\texttt{fdondelinger.work@gmail.com} \\
	  \And
	Mauricio A. Alvarez \\      
	Department of Computer Science\\
	The University of Manchester\\
	  \texttt{mauricio.alvarezlopez@manchester.ac.uk} \\
	  \And
	    Dennis Wang \\
	  Singapore Institute for Clinical Sciences (SICS) \\
        Bioinformatics Institute (BII) \\
        Agency for Science, Technology and Research (A*STAR) \\
	  \texttt{dennis\_wang@sics.a-star.edu.sg} \\
}

\renewcommand{\shorttitle}{\textit{arXiv} Template}

\hypersetup{
    pdftitle={Machine learning prediction of epigenetic age from longitudinal DNA methylation profiles},
    pdfsubject={},
    pdfauthor={Arthur Leroy, Ai Ling Teh, Frank Dondelinger, Mauricio A. Alvarez, Dennis Wang},
    pdfkeywords={DNA methylation, longitudinal data, multi-mean Gaussian processes},
    colorlinks = true,
    urlcolor = blue,
    linkcolor = orange,
    citecolor = orange
}

\begin{document}
\maketitle

\begin{abstract}
Interrogating the evolution of biological changes at early stages of life requires longitudinal profiling of molecules, such as DNA methylation, which can be challenging with children. We introduce a probabilistic and longitudinal machine learning framework based on \emph{multi-mean Gaussian processes} (GPs), accounting for individual and gene correlations across time. This method provides future predictions of DNA methylation status at different individual ages while accounting for uncertainty. Our model is trained on a birth cohort of children with methylation profiled at ages 0-4, and we demonstrated that the status of methylation sites for each child can be accurately predicted at ages 5-7. We show that methylation profiles predicted by \emph{multi-mean GPs} can be used to estimate other phenotypes, such as epigenetic age, and enable comparison to other health measures of interest. This approach encourages epigenetic studies to move towards longitudinal design for investigating epigenetic changes during development, ageing and disease progression. 
\end{abstract}


\section{Introduction}
\label{sec:intro}

%\paragraph{Context.}
Longitudinal molecular profiling is pivotal to advancing ageing research by providing critical insights into development, growth and, ultimately, frailty. By integrating a multi-omics approach with clinical data, such studies capture the dynamic nature of diseases and reveal the underlying molecular signatures of ageing and disease processes. 
Previous studies have demonstrated the power of longitudinal molecular profiling in tracking tumour evolution and identifying therapeutic targets in cancer patients \cite{Zhou2019,McMahon2017}. 
More recent investigation on population cohorts showcased how longitudinal molecular profiling aided in predicting treatment outcomes and tailoring interventions for patients with chronic diseases \cite{Li2021longprofile,Shi2022cardiovascular}. 
However, when health outcomes are recorded, particularly in children, there could be a lack of biosamples available to profile the molecular changes occurring at that time. The insufficient sampling of readouts may result in underpowered studies \citep{Guo2013samplesize}.

A particular type of molecular profile, DNA methylation, can reflect the cumulative effects of both genetic and environmental exposures, making them ideal candidates for studying long-term health outcomes. Cross-sectional studies often consider measures on a single time point, ignoring other longitudinal information that may be available. By assessing DNA methylation at multiple time points, researchers can investigate how epigenetic modifications dynamically respond to environmental stimuli, ageing, and disease progression. Researchers can estimate the epigenetic age as a function of methylation values from different signature CpGs set \citep{horvath2013dna, Pelegi2021methylclock, levine2018epigenetic, McEwen2020PedBE}.
Moreover, DNA methylation alterations have been associated with a wide range of diseases, including cancer \citep{Langevin2012necksquamous}, cardiovascular disease \citep{Shi2022cardiovascular}, diabetes \citep{Kim2021diabetes}, and psychiatric conditions \citep{Jovanova2018depression,depressioncpgs}.
Longitudinal studies incorporating DNA methylation profiles can provide insights into biological ageing processes and disease mechanisms, but they need to be repeatedly measured to capture the most important changes.

To date, there have only been algorithms to predict additional methylation sites at the same time point as the training data \citep{pretimeth2020,predictmeth2015},  
Clinicians and researchers aiming to examine methylation status at other time points when health outcomes are measured would not be able to do so if biosamples were not collected. Here, we aspire to extend the traditional cross-sectional DNA methylation analyses with a broader longitudinal approach to modelling DNA methylation across the early lifespan of children in a multi-ethnic population cohort. We demonstrate that DNA methylation at CpG sites used by epigenetic clocks can be predicted using DNA methylation patterns measured years earlier. We then correlate routine health measures with epigenetic age estimates from predicted methylation profiles.

\section{Results}
\label{sec:res}

\begin{figure}
	\centering
        \includegraphics[width=\textwidth]{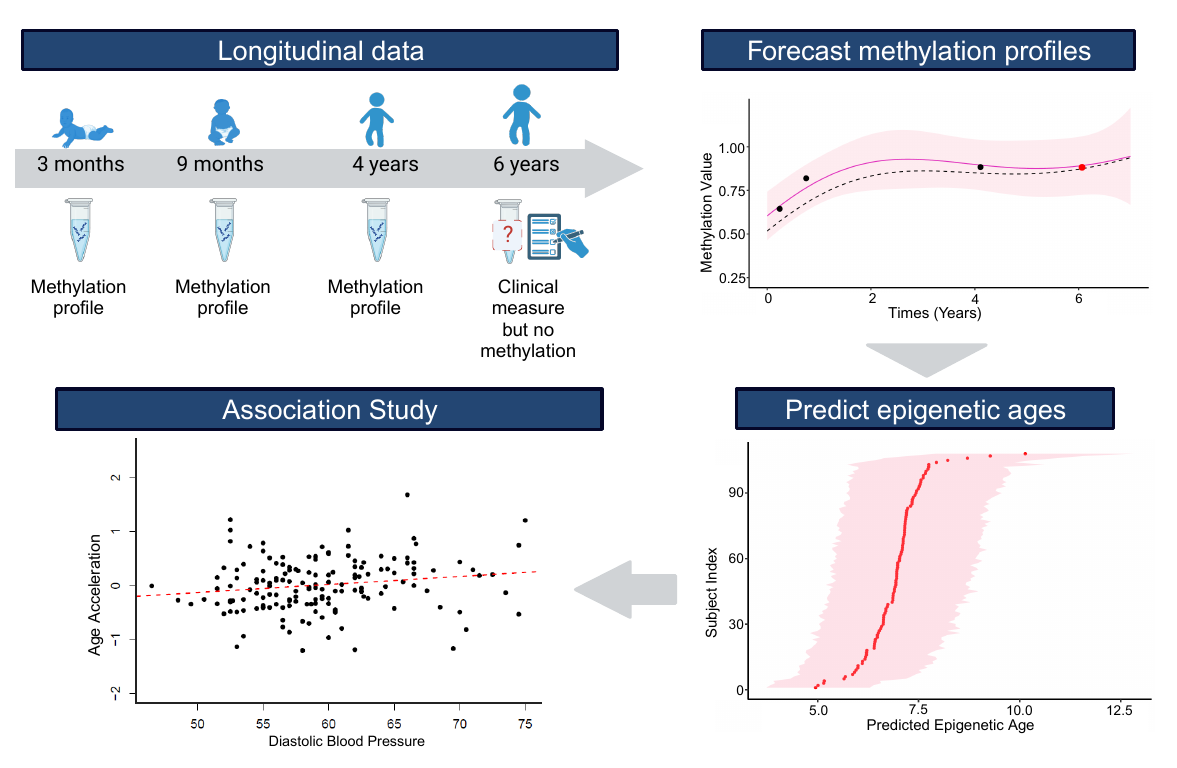}
	\caption{Schematic overview of predicting methylation profiles and its application. Methylation values were not collected when there were clinical measures. The missing methylation values at CpG sites needed for computing epigenetic age can be predicted using longitudinal data collected at early time points using multi-mean Gaussian processes. Epigenetic age computed from predicted methylation values is then used to perform the association study with clinical measure (i.e., diastolic blood pressure).}
	\label{fig:summary_fig}
\end{figure}

To demonstrate our framework and evaluate its performance, the present section illustrates the forecasting of methylation profiles and their utilisation in predicting the epigenetic age. The genomics data and phenotypes presented throughout come from the GUSTO birth cohort\citep{soh2014gusto}. From longitudinal omics data, through the \emph{multi-mean Gaussian processes} algorithm to forecast methylation values and obtain accurate epigenetic age predictions, the framework allows researchers to investigate the effect of DNA methylation at a future time and measure its association with various health outcomes \Cref{fig:summary_fig}. 

\subsection{Modelling of longitudinal methylation with uncertainty}

To study multiple DNA methylation time series simultaneously and forecast their future values, we developed a tailored machine-learning framework, based on recent developments in multi-task GPs \citep{leroy2022magma}.
The pivotal advancement comes from sharing information across all observed individuals and CpG sites by leveraging multiple adaptive mean processes, hence the name \emph{multi-mean GPs}. 

\begin{figure}
	\centering
        \includegraphics[width = 0.49\textwidth]{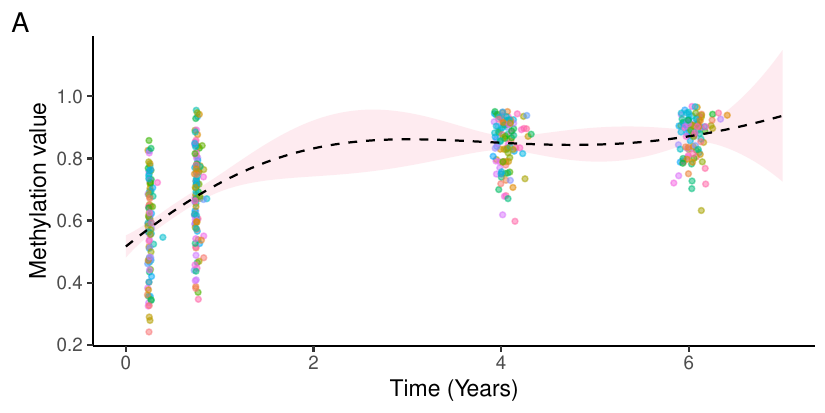}
        \includegraphics[width = 0.49\textwidth]{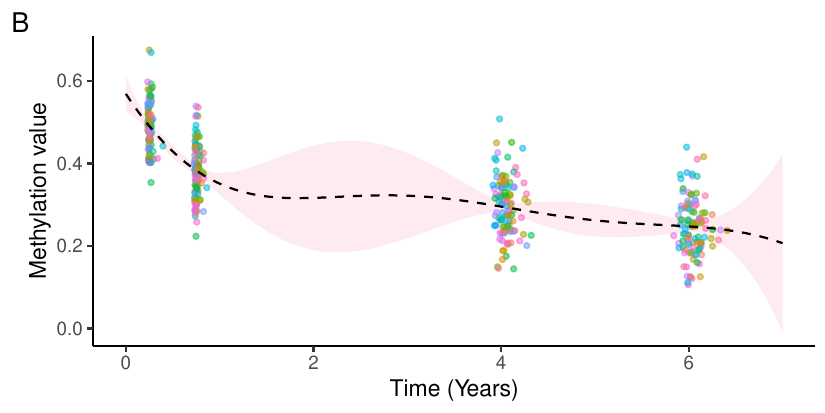}
        \includegraphics[width = 0.49\textwidth]{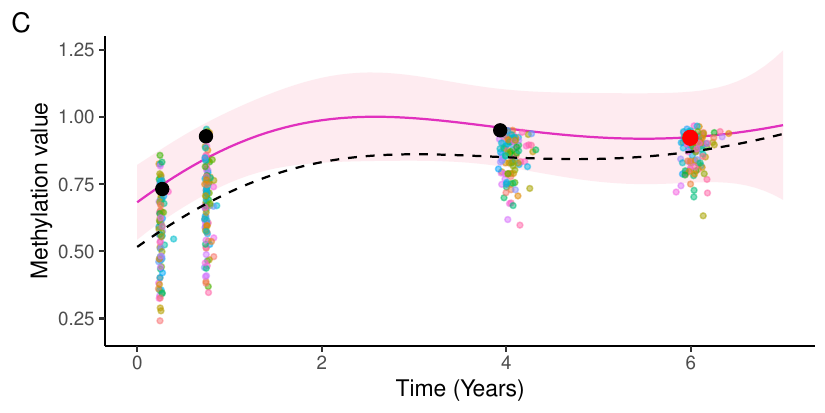}
        \includegraphics[width = 0.49\textwidth]{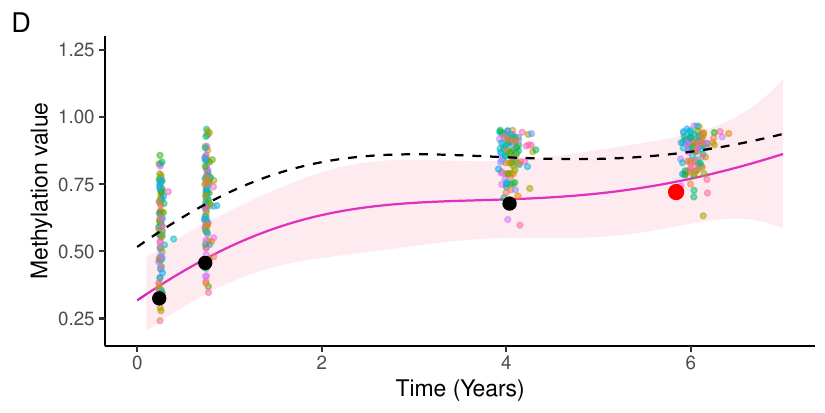}
	\caption{CpG-specific mean processes and individual-specific predictions. CpG-specific mean processes (dashed line) with associated 95\% credible interval (pink band) differs between two illustrative CpGs: cg00609333 (\textbf{A}) and cg06430061 (\textbf{B}). Multi-mean GPs prediction curve (pink) with associated 95\% credible intervals (pink band) for two illustrative individuals (\textbf{C \& D}). The dashed line represents the mean curve from the CpG-specific mean process. Observed points are coloured in black, while the predicted point is in red. Background points correspond to the training observations coloured by individuals.}
	\label{fig:mean_processes}
\end{figure}

For each CpG, the algorithm recovers the average trend of methylation values change over time (\Cref{fig:mean_processes}A and B). 
The mean estimates and their associated 95\% credible interval are estimated from data of each CpG across time points. The two examples we highlight show that the longitudinal trend and associated uncertainty of mean processes can largely differ from one CpG to another. Such adaptive behaviour is essential when it comes to predicting unobserved data points accurately from a handful of observed points.
We notice that the estimated uncertainty increases in regions lacking observations (95\% credible interval widens from 0.02 at 9 months to 0.2 around 2 years), which is an intuitive and expected property of GP-based methods. 
Conversely, the credible interval around the mean curve becomes narrow near locations with many data points, which is also expected, as abundant information is available there, leading to confident estimates.  

Although surprising at first sight, it is reasonable that the uncertainty for one individual does not cover the full range of observed methylation values in the cohort, which can be widespread across individuals (\Cref{fig:mean_processes} A).
Although we have a high variance in methylation values from one individual to another, the estimation of their shared mean will still be reliable or have low uncertainty when the number of data points increases. 

\begin{figure}
        \centering
        \includegraphics[keepaspectratio, width = \textwidth, height = 0.4\textheight]{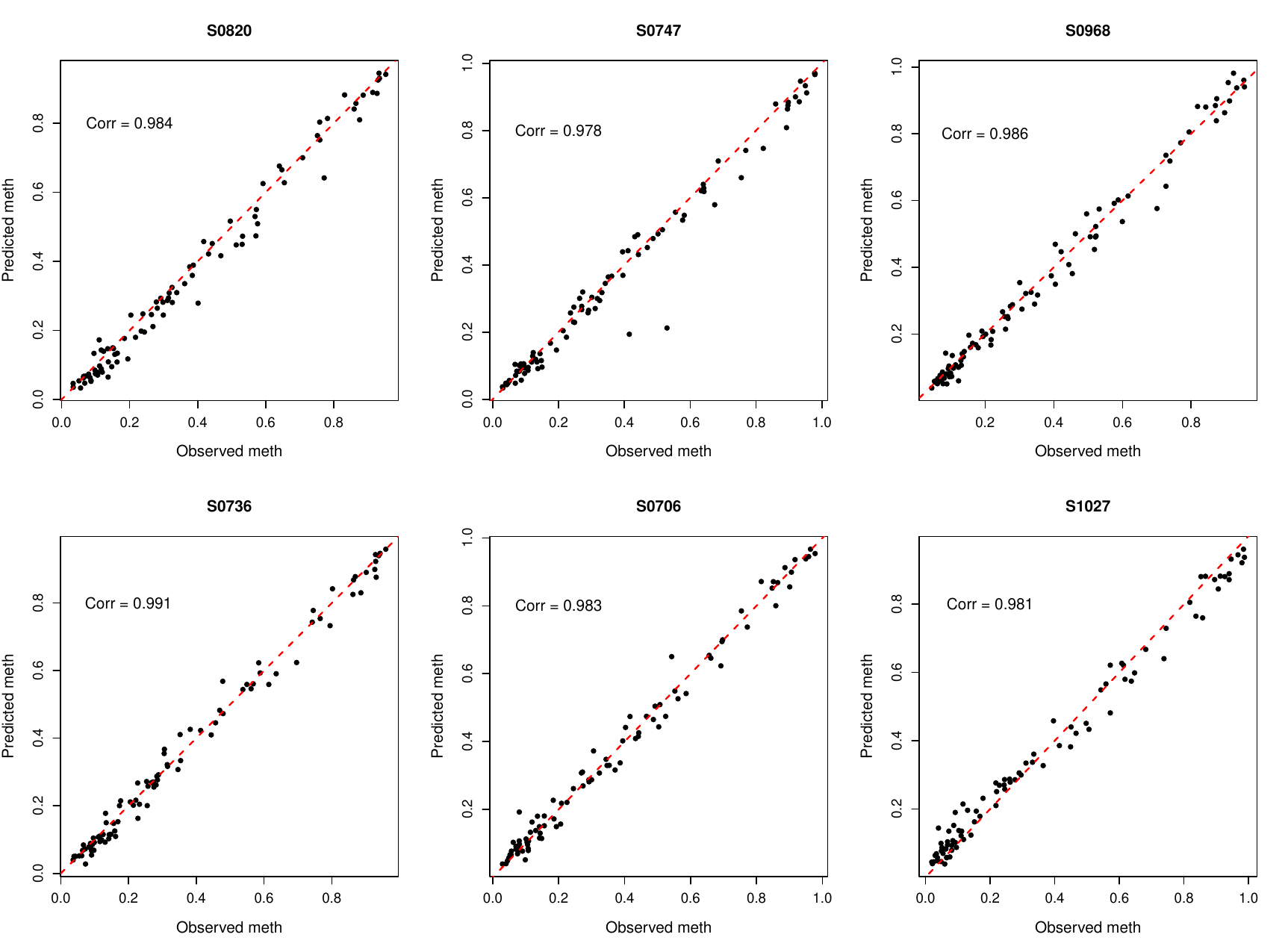}
        \caption{Examples of predicted methylation values versus measured methylation values for the 91 CpGs in PedBE clock using 6 individuals from the testing set as examples. The predicted methylation value is plotted on the y-axis, and the observed methylation is plotted on the x-axis. Each dot represents a CpG in the PedBE epigenetic clock signature. The red dotted line represents the x=y line.}
        \label{fig:pred_vs_actual_pedbe}    
\end{figure}

\subsection{Multi-mean GPs models can predict methylation status 2 years into the future}

Once mean processes have been computed for all CpGs, we aimed to forecast CpG status for all individuals at specific timepoints. We predicted methylation status at the 6-year time point based on observations at 3, 9 and 48 months. In \Cref{fig:mean_processes}, we provide an illustration of the results obtained from the prediction step of the \emph{multi-mean GPs} algorithm. By integrating out the CpG-specific mean processes, as detailed in Leroy et al. \cite{leroy2022magma} Proposition 5, it is straightforward to compute the Gaussian posterior distribution of any individual time series in closed form.
As illustrated with CpGs \emph{cg00609333} and \emph{cg06430061}, we show how the mean process acts as a common trend across all individuals, representing the general evolution pattern for this CpG (\cref{fig:mean_processes} C and D). The prediction for each individual deviates from its prior mean by adapting its trajectory to the observed data points from the individual of interest, which provides individual-specific information to enhance the accuracy of the predicted data point.

Expanding the evaluation from single CpGs, let us evaluate the proposed method's performances over 188 testing individuals and their corresponding methylation profiles of 368 and 91 CpGs for skin\&blood \citep{horvath2018skinblood} and PedBE epigenetic clock datasets \citep{McEwen2020PedBE}, respectively.
As multi-mean GPs provide a full probability distribution for predictions at 6 years, several aspects must be considered to assess their quality. 
We illustrate for 6 individuals in \Cref{fig:pred_vs_actual_pedbe} the correlation between the predicted and true methylation values for the 91 CpGs contained in PedBE's signature. We also reported correlations between predicted and observed methylation (mean Pearson=0.99; Spearman=0.98).  
We also checked the correlation of predicted and observed methylation values for each CpG (\Cref{fig:pred_vs_actual_pedbe_cpg}). We also included the overall prediction correlations for all CpGs, coloured by CpG, in \Cref{fig:true_pred_comparison}.

The most compelling advantage of GP-based methods comes from the uncertainty quantification associated with our predictions. To underline this valuable feature, we provided a visualisation of errors, computed as the difference between predicted and observed values for all individual CpGs (\Cref{fig:prediction_diff}).
The majority ($\approx 80\%$) of the errors remain within a $5$\% range ($\pm 0.05$), and roughly $95$\% of the probes present less than 10\% methylation differences ($\pm 0.10$). 
In addition, we can observe that the vast majority of errors were expected and adequately quantified by the algorithm (as we anticipated that 95\% of the data points are within the credible intervals.

%In this sense, \Cref{fig:prediction_diff} provides an alternative visualisation of errors (black dots) relative to our predictions (purple vertical line), for which we additionally displayed the 95\% credible intervals (pink area) sorted by increasing uncertainty. 
%This graph indicates that, in our predictions, we anticipated that 95\% of the black dots should be contained within the pink region. 
% One can observe that this statement is accurate for both clocks, meaning that the vast majority of errors were anticipated and adequately quantified by the algorithm. 

\begin{table}
\centering
    \caption{Performance as measured by root-mean-squared-error (RMSE) and credible interval $CIC_{95}$ for predictions of methylation values at 6 years for 188 testing individuals across all CpGs involved in PedBe and Horvath skin and blood clocks.}
\begin{tabular}{cc|cc|c|c|}
\cline{3-6}
\multicolumn{1}{l}{}          & \multicolumn{1}{l|}{} & \multicolumn{2}{c|}{multi-mean GPs}     & Individual mean & CpG mean      \\ \cline{2-2}
\multicolumn{1}{c|}{}         & Number of CpGs        & RMSE                   & $CIC_{95}$   & RMSE            & RMSE          \\ \hline
\multicolumn{1}{|c|}{skin\&blood} & 383                   & \textbf{0.046 (0.086)} & 91.90 (27.30) & 0.068 (0.112)   & 0.062 (0.104) \\
\multicolumn{1}{|c|}{PedBE}   & 94                    & \textbf{0.043 (0.068)} & 92.99 (26.90) & 0.115 (0.151)   & 0.089 (0.116) \\ \hline
\end{tabular}
 \label{tab:eval_pred}
\end{table}

%In order to provide numerical evidence to support the previous graphical evaluation. 

We can confirm that, as expected, the ratios of empirical errors contained within the 95\% credible interval are, on average, 92.99\% for skin\&blood and 91.65\% for PedBE, close to the theoretical value \Cref{tab:eval_pred}.

To evaluate the mean predictive performances, we also compared our results to natural baselines. First, we called \emph{individual mean}, the average of methylation values at 3, 9, and 48 months as an estimation of the individual typical values. 
Conversely, we also proposed the \emph{CpG-mean} as another estimator, computed as the average of methylation values at 6 years from all other individuals. Those estimators use information that multi-mean GPs somewhat combine to offer predictions leveraging both the mean trend at 6 years and the individual-specific pattern from previous measurements, resulting in more accurate predictions and a lower RMSE overall (\Cref{tab:eval_pred}). 

\subsection{Epigenetic age can be estimated from predicted methylation profiles}
\label{sec:eval_pred_age}

%From a methodological point of view, predicting hundreds of sparsely observed and correlated time series can be a tedious task. 
In addition to forecasting CpG status, considerable utility can be derived from investigating the relationship between methylation profiles and health outcomes from an individual. We explored how the DNA methylation time series can be leveraged to anticipate the evolution of biological outcomes, even in the long term, while propagating the uncertainty associated with our predictions. Here, we exemplify this by looking at the prediction of epigenetic age.

\begin{figure}
	\centering 
        \includegraphics[width = \textwidth]{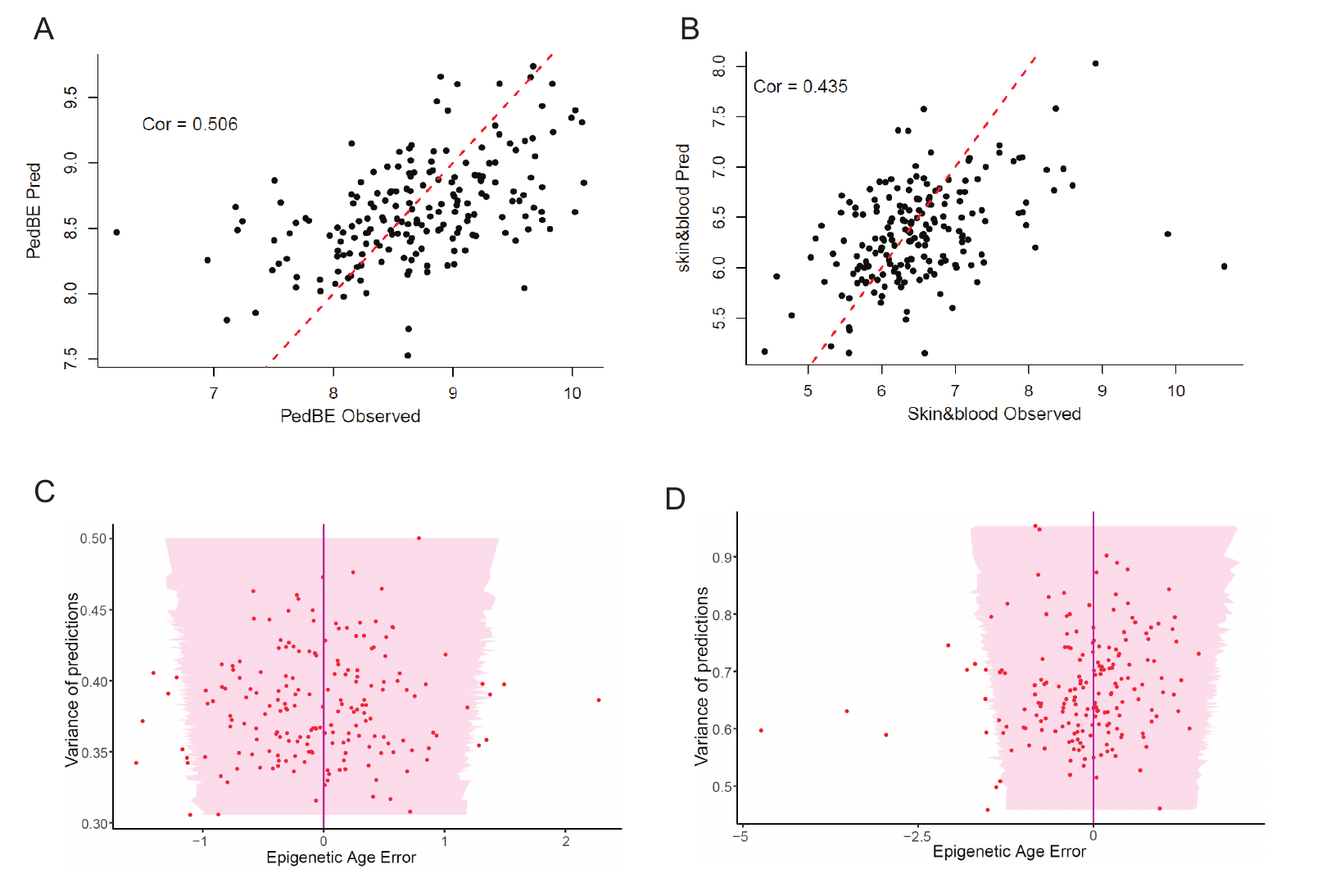}
        \caption{Epigenetic age computed from predicted methylation values versus observed methylation values and the variance of predictions for 188 testing samples. Epigenetic ages computed from predicted methylation values are plotted against epigenetic age computed from observed methylation values for PedBE clock (\textbf{A}) and skin\&blood clock (\textbf{B}). The variance of the epigenetic age prediction (uncertainty quantification of the epigenetic age predictions) is plotted against errors (difference in epigenetic age from using predicted methylation and observed methylation value) for 188 individuals for PedBE (\textbf{C}) and skin\&blood (\textbf{D}) respectively. For each individual (sorted by increasing uncertainty on the y-axis), the predicted mean ages using  PedBE and skin\&blood clocks on \emph{predicted methylation values at 6 years} are used as a reference and displayed as a purple line; the pink region corresponds to the associated 95\% credible intervals; each red dot corresponds to the epigenetic age computed using \emph{true observed methylation values.}}
        \label{fig:epigenetic_age_error}
\end{figure}

The epigenetic age is computed from methylation values following skin\&blood \citep{horvath2018skinblood} and PedBE \citep{McEwen2020PedBE} clocks.
An example of these estimations is depicted in \Cref{fig:illu_age_pred} for an illustrative individual. 
Our predictions are displayed as probability distributions, and we can see they accurately recover the true epigenetic age (red line) while accounting for the uncertainty propagated from the underlying methylation forecasts.
We observed that the epigenetic age computed following the skin\& blood clock is generally closer to the chronological age than with the PedBE clock. 
Indeed, epigenetic age estimated from PedBE is higher than chronological age for the majority of samples. 
Let us mention that we retrieve such behaviour numerically in \Cref{tab:eval_age}, and the dedicated \Cref{fig:PedBE_bias} illustrates more clearly the shift between epigenetic and chronological age, depending on the considered clock.
This trend has previously been observed in other studies \citep{Felixpedbeusecase,PopovicPedBE}.
Nonetheless, the epigenetic age computed from predicted methylation values and observed methylation values are well correlated (\Cref{fig:epigenetic_age_error}, A and B).

We evaluated the predictive performance by displaying absolute errors with the associated uncertainty of the epigenetic age predictions for the 188 testing individuals (\Cref{fig:epigenetic_age_error}, C and D). 
Similarly to the methylation time series predictions, we remark that the range of empirical errors of the epigenetic age (red points) remains adequately recovered by the pink credible intervals. 
This visual intuition is confirmed in \Cref{tab:eval_age}, which reports the performance metrics regarding mean accuracy and empirical uncertainty coverage.
Once again, the multi-mean GPs approach leads to credible interval coverages that remain close to the theoretical value of 95\%. 
This property is especially reassuring from a practical point of view as it indicates that high uncertainty should be dealt with additional caution when conducting inference. 

\subsection{Predicted age acceleration is associated with adolescent health outcomes}

\begin{figure}
	\centering
        \includegraphics[keepaspectratio, width = \textwidth, height = 0.4\textheight]{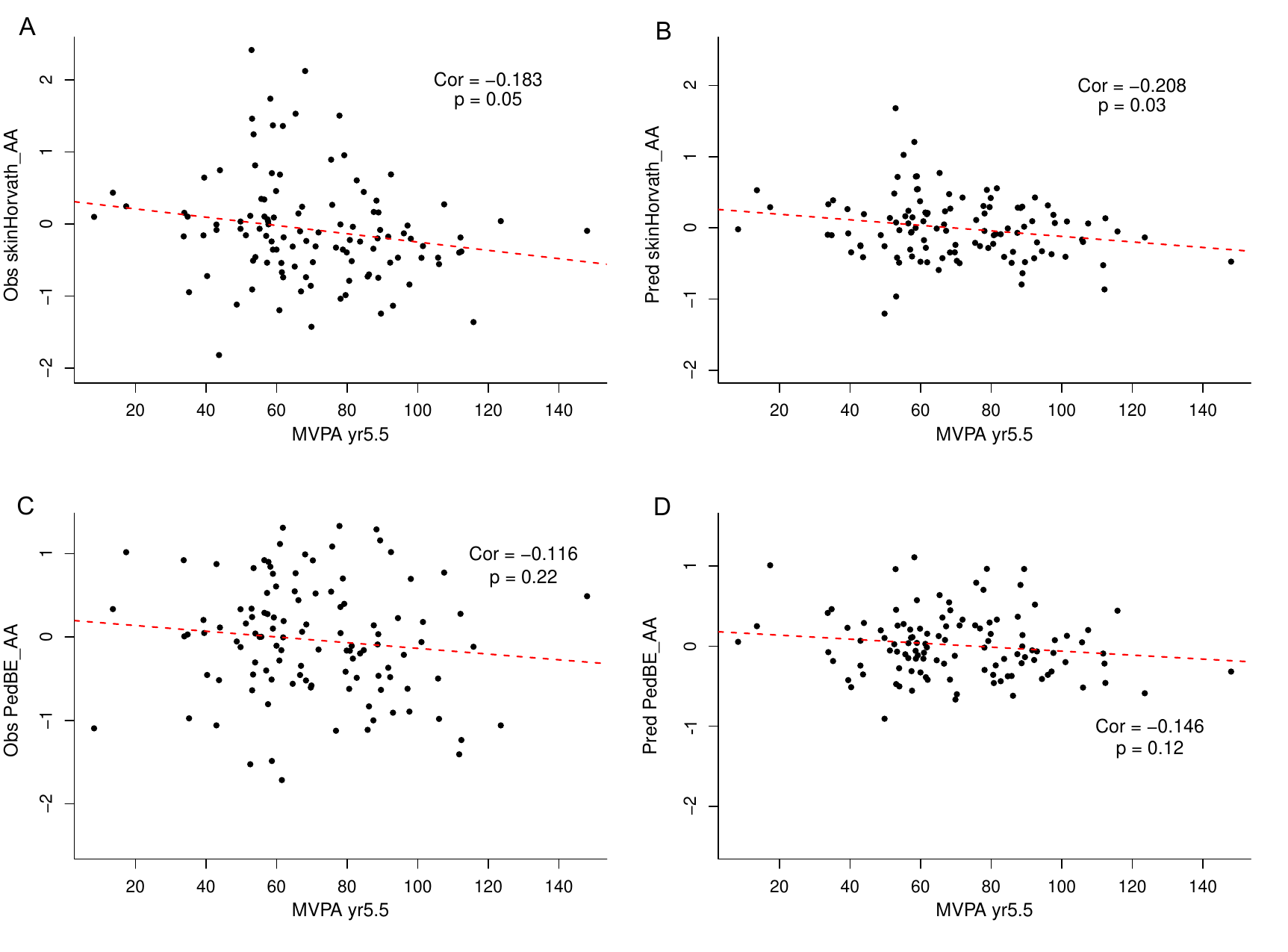}
	\caption{Scatter plot of age acceleration(AA) plotted against moderate to vigorous physical activity (MVPA) measured at age 5.5 years. Each dot represents a subject in the testing set. The red dotted line is the regression line.}
	\label{fig:mvpa_assoc}
\end{figure}
Using the epigenetic age computed from the predicted methylation values, we have also computed the age acceleration (AA), defined as the residuals of regressing epigenetic age and chronological age\citep{Pelegi2021methylclock} for the subjects. 
We then performed association tests with clinical variables such as \emph{moderate to vigorous physical activity} (MVPA) and measured blood pressure.
The results obtained from both observed and predicted methylation values are similar, though the significant levels are different (\Cref{fig:mvpa_assoc}). 
From this figure, we can notice that a higher MVPA measured at age 5.5 years old is associated with lower AA for skin\&blood (p$_{obs}$ = 0.05, p$_{pre}$ = 0.03).
Regarding PedBE, despite not being significant, the trend of age acceleration and MVPA remains similar.
In addition, we also observed that higher diastolic blood pressure is associated with higher AA at 6 years from predicted methylation values (PedBE: p$_{pre}$ = 0.02, skin\&blood: p$_{pre}$ = 0.03). 
Although these subjects did not present significant associations between measured AA and systolic blood pressure, it remains 
noteworthy that both associations present similar trends (\Cref{fig:dbp_assoc}). 
Overall, the results from predicted methylation values appear as a faithful reflection of methylation status from early time points, which always constitutes valuable guidance for practitioners.

\section{Discussion}
\label{sec:discussion}

 We presented the first-ever attempt to forecast DNA methylation profiles of human individuals longitudinally.
 Our proposed approach is able to predict the methylation value with less than 10\% differences between the observed and predicted value in about 95\% of the data.
 Note that the algorithm can produce predictions at any time point, allowing for interpolation and filling in missing data. 
 The choice of the 6-year validation time-point was made to demonstrate the ability to predict future methylation states from an existing time series. 
 More importantly, we demonstrated that the uncertainty associated with predictions is well-calibrated, which informs downstream users on the confidence they should grant to each forecast.
 We observed that higher MVPA is associated with lower AA for children. 
 A recent study has also reported a similar correlation for adult subjects \citep{fox2023mvpaepiage}.
 Let us point out that the trends of MVPA and age acceleration look consistent across age groups as well.  

Throughout the present paper, we limited the predictions to the subset of CpGs required to estimate epigenetic age. 
Nevertheless, our algorithm can be scaled (the computational complexity is linear in the number of individuals and CpGs) to make predictions for all potential methylation sites on the human genome. 
This would enable the identification and validation of epigenetic signatures associated with health conditions, such as obesity \citep{obesitycpgs,BMIsignature} and mental health \citep{depressioncpgs}.
Such advances would contribute vastly to the research community conducting epigenetic studies.

In this study, we were able to make accurate predictions of methylation status around age 6 based on methylation profiles collected from three early time points.
There is potential for our method to perform better if we trained on data recorded more often during the period of measures.
Nonetheless, the proposed algorithm can easily include more data, possibly coming from different sources or studies, to train mean processes on a broader range of ages, resulting in more accurate predictions.
Optimising the number of data points needed to forecast the methylation profile X years into the future will help researchers plan their studies and determine when to collect data from their subjects. 
Potential future studies could profile epigenetic ageing processes throughout the whole lifespan of an individual, from infancy to elderly, without incurring unnecessary costs or burdens for the participant.

\section{Online Methods}
\label{sec:methods}

This section describes the studied cohort, the variables of interest, and how the data was collected.
We also describe step-by-step, including technical details, how to derive probabilistic epigenetic age predictions from methylation time series. 

\subsection{Description of the child cohort}
\label{sec:gusto_description}

The subjects in this analysis are participants of the \emph{Growing Up in Singapore Towards healthy Outcomes} (GUSTO) birth cohort \citep{soh2014gusto}. 
Briefly, this multi-ethnic study gathers about 1400 mother-child pairs with dense phenotypes.
Mothers were recruited at two participating hospitals during the first trimester of pregnancy and followed over time through regular clinic/home visits. 
The children of these women are also followed over time after birth. 
To conduct this analysis, we selected 110 training subjects for whom we had access to longitudinal methylation data profiles at 3, 9, 48 and 72 months. 
We then validated our model using another set of 188  testing subjects with methylation data at the time points mentioned. 
The basic demographics of these subjects are shown in \Cref{tab:demo_table}. 
We refer to those timestamps in the sequel as 3, 9, 48, and 72 months, although the exact time of data collection differs from one individual to another.
All computations were performed using the exact date of data collection. 
Since the 72-month data point can sometimes be observed at 70 months for one individual and at 80 months for another, we thus accounted for the possible influence of such time lags in measurements. 
For example, this variance in data collection times is well-illustrated in \Cref{fig:mean_processes}, where points correspond to raw training data coloured by individuals. 

\subsection{DNA methylation data}

Buccal swab samples were collected from the participants by research staff using Isohelix swabs (SK-2S) during clinic visits and stored at -80\textdegree C freezer.
DNA was then extracted from buccal swabs using IsoHelix DNA Extreme Kit (0-4 years samples) and Qiasymphony SP (6 years samples) following the standard protocol recommended by the company.

Extracted DNA samples were sent for DNA methylation profiling using the Infinium MethylationEPIC BeadChip array (EPIC850k). DNA methylation profiling was done as per Illumina standard protocol.
Data quality control (QC) and preprocessing were performed in R.
The raw .idat files were read and processed using the \emph{minfi} package (version 1.42.0)\cite{Aryee2014}. 
The methylation value for a CpG was denoted "NA" when the detection p-value > 0.0001 or NBeads < 3. 
A CpG was removed from the analysis if more than 5\% of the samples failed the QC criteria.
Probes that contain SNP at the CpG position or single base pair and cross-hybridization probes that mapped to multiple positions were also removed.

\subsection{Clinical variable collection}

Clinical variables such as diastolic blood pressure and systolic blood pressure were collected by trained clinicians during the clinic visit. 
Anthropometric data such as weight and height were also measured during clinic visits. 
The physical activity (moderate to vigorous) measurements were collected using an accelerator. 

\subsection{Longitudinal modelling}
\label{sec:modelling}

In order to model the methylation values over time for each CpG and individual as continuous functions (i.e. time series), we proposed a framework based on Gaussian processes, which can be considered as an extension of the algorithm presented in Leroy et al. \cite{leroy2022magma}. 
We called this framework \emph{multi-mean Gaussian processes} to emphasise the strategy leveraging multiple latent mean processes to provide adaptive predictions in cases of a large number of time series presenting multiple sources of correlation. 
In the present study, we can define two separate sources of correlations in our data: the individuals and the CpGs (as we can expect time series coming from the same individual, or CpG, to present related patterns). 
Let us denote $y_i^j(t)$ the DNA methylation value associated with the $i$-th individual, the $j$-th CpG, observed at a time $t$.
From a mathematical point of view, the proposed multi-mean GPs model can be expressed as follows:
\begin{equation*}
    y_i^j(t) = \mu_0(t) + f_i(t) + g^j(t) + \epsilon_i^j(t), \ \ \forall t \in \mathcal{T}, \ \ \forall i \in \I, \ \ \forall j \in \J, 
\end{equation*}
where $\mu_0(\cdot) \sim \mathcal{GP} (m_0(\cdot), k_{\theta_0}(\cdot,\cdot))$ is a mean common process, while $f_i(\cdot) \sim \mathcal{GP} \paren{0,k_{\theta_i}(\cdot,\cdot)}$ represents an individual-specific process, and $g^j(\cdot) \sim \mathcal{GP} \paren{0, k_{\theta^j}(\cdot,\cdot)}$ a CpG-specific process.
Moreover, the error term is supposed to be $\epsilon_i^j(\cdot) \sim \mathcal{GP} (0, \sigmaij I_d)$.

Intuitively, this means that any time series $y_i^j$ is assumed to be the sum of a common mean trend ($\mu_0$), a perturbation coming from the individual ($f_i$), and another perturbation specific to the CpG ($g_j$). 
For readers familiar with mixed models in statistics, one can retrieve ideas roughly similar to the notions of \emph{fixed} and \emph{random} effects, although we are here working with continuous functions instead of vectors. 
For the sake of clarity, the graphical model, illustrating this data generative process and the modelling assumptions, is provided in \Cref{graph_model}.

\subsection{Methylation prediction}
\label{sec:meth_pred}

Leaving mathematical sophistication aside, it is possible to leverage an EM algorithm (as proposed in Leroy et al.\cite{leroy2022magma}) to estimate both the model parameters and the mean process from data. 
More specifically, the key idea of the present framework consists in computing \emph{multiple} mean processes. 
Each mean process is associated with a specific CpG by computing a posterior distribution conditioned over an adequate subset of data through Bayes' law. 
This unusual (though essential) choice allows us to derive CpG-specific mean processes, which are particularly relevant for predicting future values of the time series. 
Roughly speaking, one can think of those processes as averaged curves over all individuals, along with their associated uncertainty quantification. 
An example of two different CpG-specific mean processes is displayed in \Cref{fig:mean_processes} to illustrate the ability of multi-mean GPs to recover an adaptive trend from each subset of data accurately. 
Although intuitive and seemingly trivial, these mean processes are of utmost importance when it comes to predicting unobserved values for a particular individual. 
Whether the goal consists in predicting missing data or (possibly long-term) forecasting from a handful of points, those mean processes provide a powerful way to implicitly transfer knowledge across individuals to improve performances in these tasks.
The current implementation of the framework, based on the \emph{R} package MagmaClustR \citep{MagmaClustR}, is provided in the following repository: \url{https://github.com/ArthurLeroy/MultiMeanGP}.

In the experiments reported in \Cref{sec:res}, we considered the time series of methylation profiles for 110 training individuals observed at 368 (skin\&blood clock) and 91 (PedBE clock) CpG sites, respectively.
Each time series is specific to an individual-CpG couple and consists of measurements collected at 4 timestamps (3, 9, 48 and 72 months). 
The objective resides in predicting methylation values at 72 months for any of the 188 testing individuals who were partially observed (i.e. with observed data only at 3, 9 and 48 months or any subset of these). 
From a mathematical point-of-view, this problem is far from trivial as such long-term forecasting tasks (a 2-year gap between the last observation and the target value) from only 3 data points generally lead to unsatisfactory results.
However, leveraging the mean processes trained with the multi-mean GPs algorithm can greatly enhance performances by sharing information and uncertainty quantification across individuals.
As illustrated in the bottom panel of \Cref{fig:mean_processes}, for two different individuals, the prediction of multi-mean GPs benefits both from the mean process' ability to recover the long-term trend of this particular CpG, as well as the specific pattern coming from the predicted individual.  
The methylation forecasting step ends when we obtain a predicted value at 72 months (more specifically, at the exact timestamp of data collection around 72 months) for all 110 individuals and all 368 or 91 CpGs. 
Then, from these values, we can expect to recover estimations for the epigenetic age. 

\subsection{Estimating epigenetic age}
\label{sec:estim_age}

As previously mentioned, various methods exist nowadays to compute epigenetic age from an adequate CpG signature. 
In the present paper, we took advantage of two well-known epigenetic clocks, namely the Horvath skin\&blood  \cite{horvath2018skinblood} and the PedBE \cite{McEwen2020PedBE} clocks. 
Both methods leverage the same functional form to express the relationship between epigenetic age and methylation values through the following equation (see \citet{horvath2013dna}, Additional file 2):
\begin{equation}
\label{eq:clock}
   age = 21 \times \exp (\textbf{c}^{\intercal} \textbf{x} ) - 1,
\end{equation}
where $\textbf{x}$ represents the vector of methylation values for each CpG and $\textbf{c}$ the vector of associated coefficients.
While the formula is similar for both clocks, the subset of CpGs involved in the computation differs.
Therefore, $\textbf{x}$ and $\textbf{c}$ are 368-dimensional vectors in the Horvath skin\&blood clock and 91-dimensional vectors in the PedBE one. 
What is actually called a \emph{clock} here corresponds to the aforementioned \cref{eq:clock}, along with a vector of coefficients $\textbf{c}$, which differs depending on the method used.
The vector of methylation values $\textbf{x}$ is generally observed, thus allowing direct computation of epigenetic age. 
In our study, we instead used predictions obtained in \Cref{sec:meth_pred} as a surrogate to the true observations $\textbf{x}$.
As our predictions are defined as Gaussian distributions, we cannot extract a single vector $\textbf{x}$ without losing information about uncertainty quantification. 
In order to propagate the probabilistic nature of our forecasting through \cref{eq:clock}, we merely generated a large number of samples (10,000 samples for each clock) from the predictive GP distributions associated with each CpG. 
Hence, each sample corresponded to a single vector $\textbf{x}$ for which estimating the epigenetic age through \cref{eq:clock} is trivial. 
Finally, the resulting set of 10,000 epigenetic age estimations provided an empirical distribution accounting for the full uncertainty in our predictions. 
Although we proposed 10,000 samples as a sufficient number to obtain accurate empirical estimates, it is straightforward to increase this number arbitrarily in our implementation to make those results as close as desired to the theoretical distribution. 
An illustration of those empirical distributions for estimated epigenetic age is provided in \Cref{sec:eval_pred_age}.
In order to evaluate the accuracy of our predictions, we compared them to the actual epigenetic age (computed from the observed values at 72 months) and to the true age of data collection for all individuals (\Cref{fig:PedBE_bias}).
Extensive evidence is provided in \Cref{sec:eval_pred_age} through various computations of error metrics and visualisation of performances.  

\subsection{Evaluation metrics}
For clarity, let us recall that $N$ denotes the number of individuals, $T_i$ the number of time points observed for the $i$-th individual, whereas $y_{obs}$ and $y_{pred}$ represent the vectors of observed and predicted methylation values, respectively.
Formally, we define the Root Mean Squared Error (RMSE) in the subsequent experiments as follows:
\begin{equation*}
	\sqrt{\dfrac{1}{N \times T_i} \sum\limits_{i = 1}^{N}  \sum\limits_{t = 1}^{T_i} \paren{y_i^{obs}(t) - y_i^{pred}(t) }^2}.
\end{equation*}
Moreover, an additional measure of uncertainty quantification \citep{leroy2022magma} is used to evaluate whether the observations belong to the predicted credible interval as expected.
Namely, the $CI_{95}$ coverage ($CIC_{95}$) is defined as:
\begin{equation*}
	100 \times \dfrac{1}{N} \sum\limits_{i = 1}^{N} \ \mathds{1}_{ \{ y_i^{obs} \in \ CI_{95} \}},
\end{equation*}
\noindent where $CI_{95}$ represents the 95\% credible interval computed from the predictive Gaussian distribution.
When interpreting this metric, the closer to the theoretical value of 95\%, the better. 

\subsection{Statistical Analysis}
\label{sec:analysis}
All statistical analyses were performed in RStudio \citep{rstudio} (version 2022.12.0 Build 353). 
Simple linear regression was performed to test the association between age acceleration and clinical relevance variables. 
Two subjects with observed AA above the 99th percentile were excluded from the analyses.

\newpage
\bibliographystyle{unsrtnat}
\bibliography{biblio}  %%% Uncomment this line and comment out the ``thebibliography'' section below to use the external .bib file (using bibtex).

\newpage
\section{Acknowledgement}
The study is supported by the Wellcome Trust Longitudinal Population Studies Grant (217068/Z/19/Z).
We would also like to thank the National Research Foundation (NRF) under the Open Fund-Large Collaborative Grant (OF-LCG; MOH-000504) administered by the Singapore Ministry of Health’s National Medical Research Council (NMRC) and the Agency for Science, Technology and Research (A*STAR). In RIE2025, GUSTO is supported by funding from the NRF’s Human Health and Potential (HHP) Domain, under the Human Potential Programme.

We would like to thank GUSTO participants and GUSTO study group which includes Airu Chia, Allan Sheppard, Amutha Chinnadurai, Anna Magdalena Fogel, Anne Eng Neo Goh, Anne Hin Yee Chu, Anne Rifkin-Graboi, Anqi Qiu, Arijit Biswas, Bee Wah Lee, Birit Froukje Philipp Broekman , Bobby Kyungbeom Cheon, Boon Long Quah, Candida Vaz, Chai Kiat Chng, Cheryl Shufen Ngo, Choon Looi Bong, Christiani Jeyakumar Henry, Ciaran Gerard Forde, Claudia Chi, Daniel Yam Thiam Goh, Dawn Xin Ping Koh, Desiree Y. Phua, Doris Ngiuk Lan Loh, E Shyong Tai, Elaine Kwang Hsia Tham, Elaine Phaik Ling Quah, Elizabeth Huiwen Tham, Evelyn Chung Ning Law, Evelyn Xiu Ling Loo, Fabian Kok Peng Yap, Faidon Magkos, Falk Müller-Riemenschneider, George Seow Heong Yeo, Hannah Ee Juen Yong, Helen Yu Chen, Heng Hao Tan, Hong Pan, Hugo P S van Bever, Hui Min Tan, Iliana Magiati, Inez Bik Yun Wong, Ives Yubin Lim, Ivy Yee-Man Lau, Izzuddin Bin Mohd Aris, Jeannie Tay, Jeevesh Kapur, Jenny L. Richmond, Jerry Kok Yen Chan, Jia Xu, Joanna Dawn Holbrook, Joanne Su-Yin Yoong, Joao Nuno Andrade Requicha Ferreira, Johan Gunnar Eriksson, Jonathan Tze Liang Choo, Jonathan Y. Bernard, Jonathan Yinhao Huang, Joshua J. Gooley, Jun Shi Lai, Karen Mei Ling Tan, Keith M. Godfrey, Kenneth Yung Chiang Kwek, Keri McCrickerd, Kok Hian Tan, Kothandaraman Narasimhan, Krishnamoorthy Naiduvaje, Kuan Jin Lee, Leher Singh, Li Chen, Lieng Hsi Ling, Lin Lin Su, Ling-Wei Chen, Lourdes Mary Daniel, Lynette Pei-Chi Shek, Marielle V. Fortier, Mark Hanson, Mary Foong-Fong Chong, Mary Rauff, Mei Chien Chua, Melvin Khee-Shing Leow, Michael J. Meaney, Michelle Zhi Ling Kee, Min Gong, Mya Thway Tint, Navin Michael, Neerja Karnani, Ngee Lek, Oon Hoe Teoh, P. C. Wong, Paulin Tay Straughan, Peter David Gluckman, Pratibha Keshav Agarwal, Priti Mishra, Queenie Ling Jun Li, Rob Martinus van Dam, Salome A. Rebello, Sambasivam Sendhil Velan, Seang Mei Saw, See Ling Loy, Seng Bin Ang, Shang Chee Chong, Sharon Ng, Shiao-Yng Chan, Shirong Cai, Shu-E Soh, Sok Bee Lim, Stella Tsotsi, Stephen Chin-Ying Hsu , Sue-Anne Ee Shiow Toh, Suresh Anand Sadananthan, Swee Chye Quek, Varsha Gupta, Victor Samuel Rajadurai, Walter Stunkel, Wayne Cutfield, Wee Meng Han, Wei Wei Pang, Wen Lun Yuan, Yanan Zhu, Yap Seng Chong, Yin Bun Cheung, Yiong Huak Chan, Yung Seng Lee.

\newpage
\section{Supplementary}
\label{sec:supplementary}
\renewcommand{\figurename}{Fig.}
\renewcommand{\thefigure}{S\arabic{figure}}
\renewcommand{\tablename}{Table}
\renewcommand{\thetable}{S\arabic{table}}

\begin{figure}[hbt!]
\centering\includegraphics[keepaspectratio, width = \textwidth, height= 0.4\textheight]{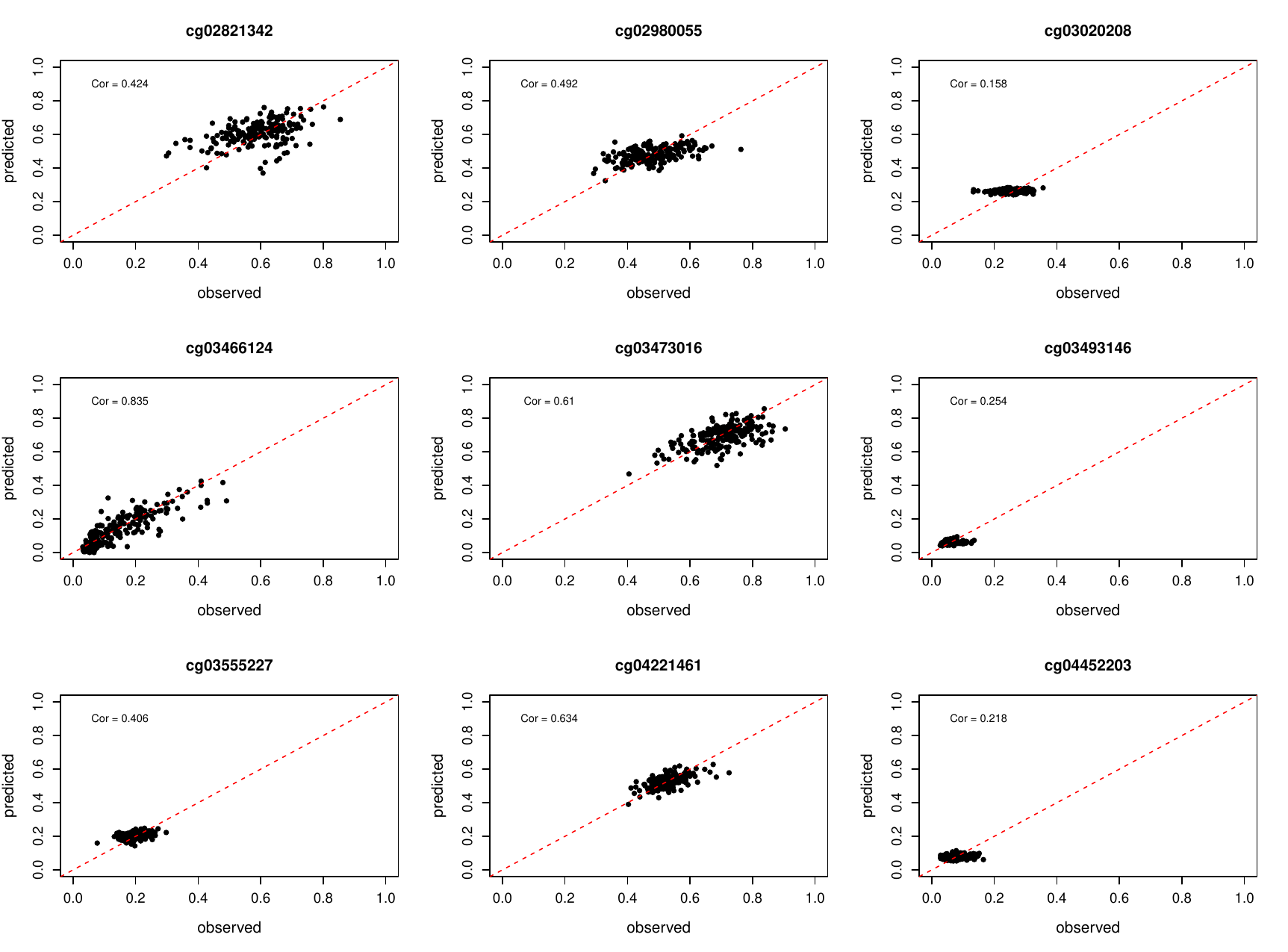}
    \caption{Example of predicted methylation values against observed methylation values for the CpGs in PedBE clock plotted for 6 illustrative CpGs.}
    \label{fig:pred_vs_actual_pedbe_cpg}
\end{figure}

\begin{figure}[hbt!]
	\centering
        \includegraphics[width = 0.49\textwidth]{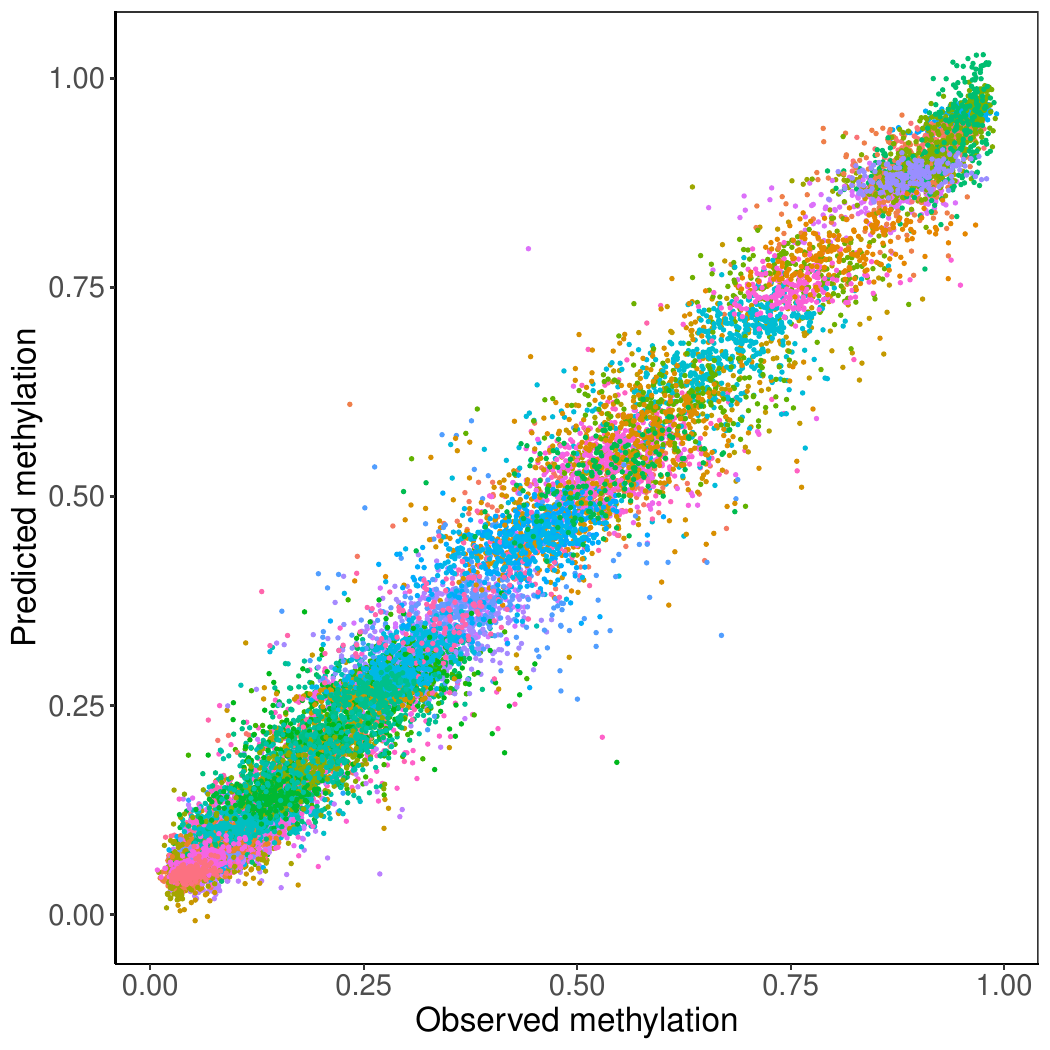}
       \includegraphics[width = 0.49\textwidth]{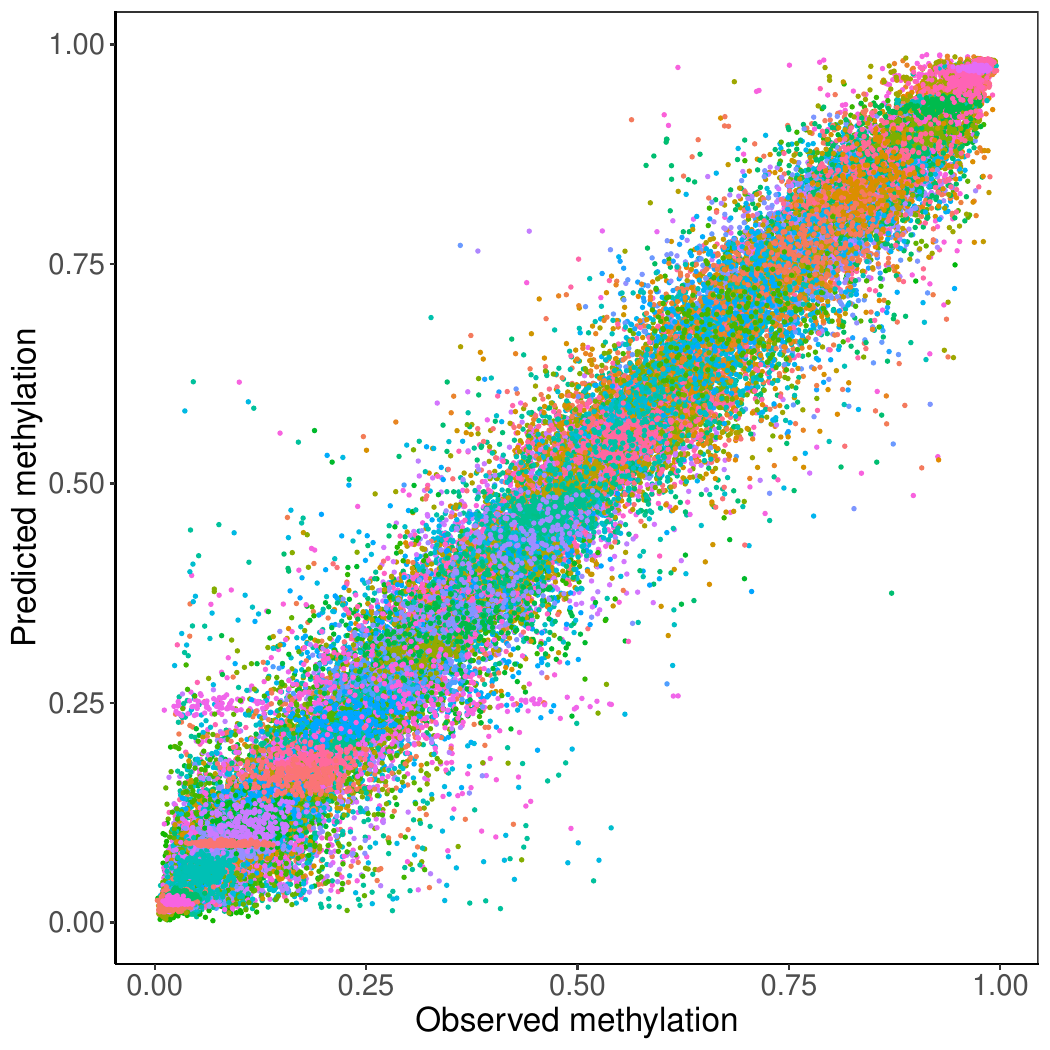}
	\caption{Mean predicted values (y-axis) plotted against observed methylation values (x-axis) for each CpG-individual couple of the testing set (188 individuals) involved in the PedBE clock (91 CpGs, \textbf{left}) and the Horvath skin\&blood clock (368 CpGs, \textbf{right}). Each colour corresponds to a specific CpG.}
	\label{fig:true_pred_comparison}
\end{figure}

\begin{figure}
	\centering
        \includegraphics[width = 0.49\textwidth]{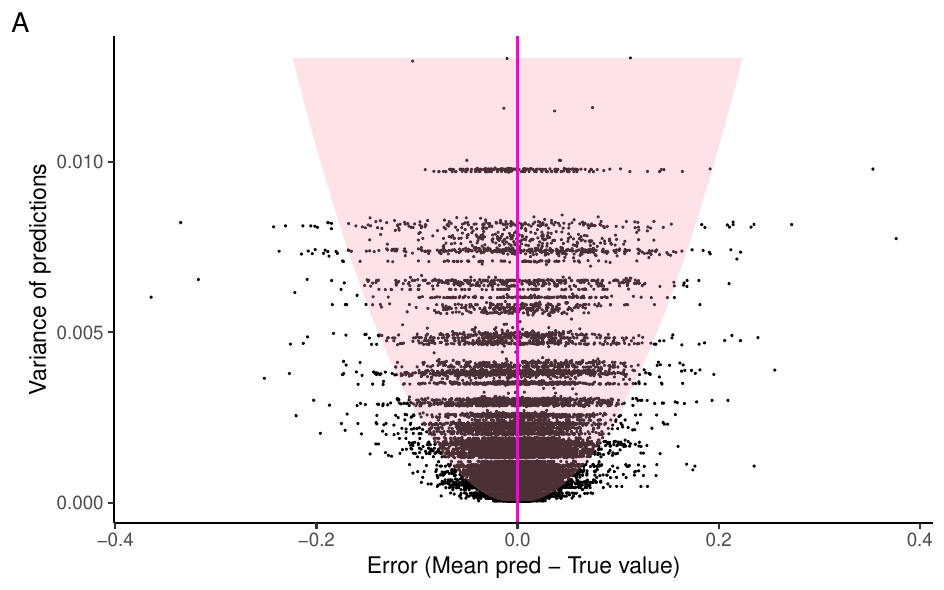}
        \includegraphics[width = 0.49\textwidth]{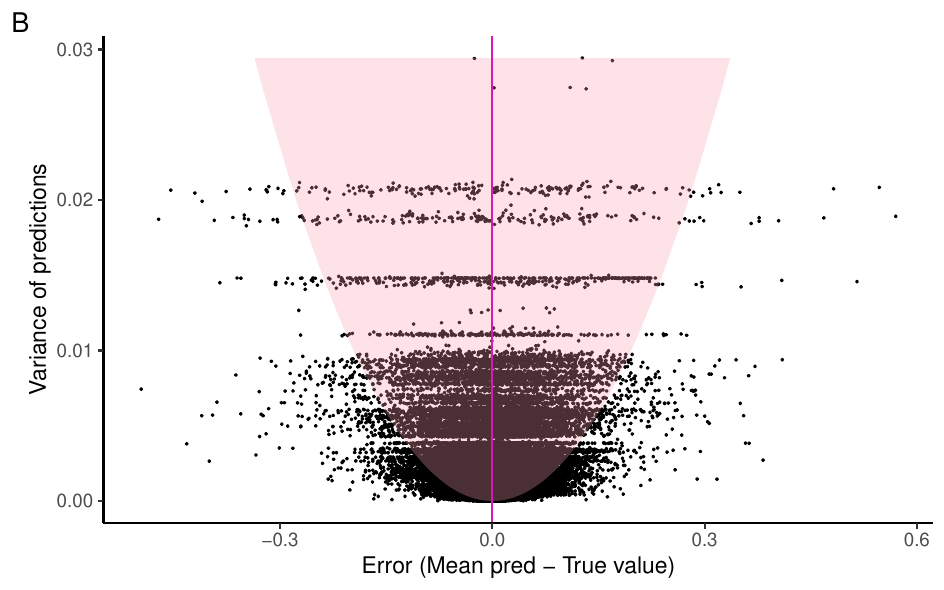}
        \caption{Distribution of prediction errors (differences between predicted and observed methylation value) sorted by predictive variance for both PedBE clock (\textbf{A}) and skin\&blood (\textbf{B}). For each individual-CpG couple of the testing set (188 individuals), each error value is displayed as a black dot, and the pink region represents the 95\% credible interval associated with the predicted mean (purple vertical line).}
        \label{fig:prediction_diff}
\end{figure}

\begin{figure}[hbt!]
	\centering
        \includegraphics[keepaspectratio, width = 0.49\textwidth, height = 0.35\textheight]{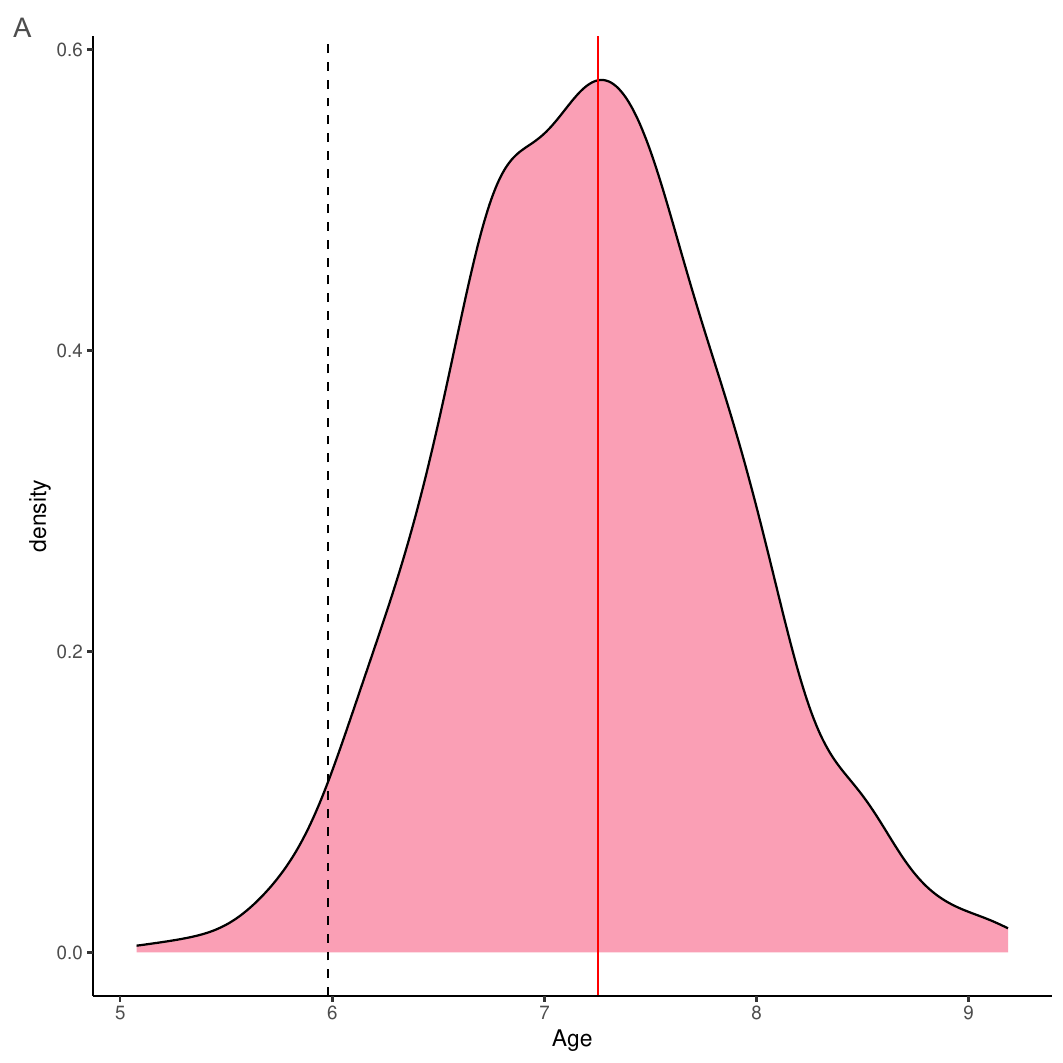}
        \includegraphics[keepaspectratio, width = 0.49\textwidth, height = 0.35\textheight]{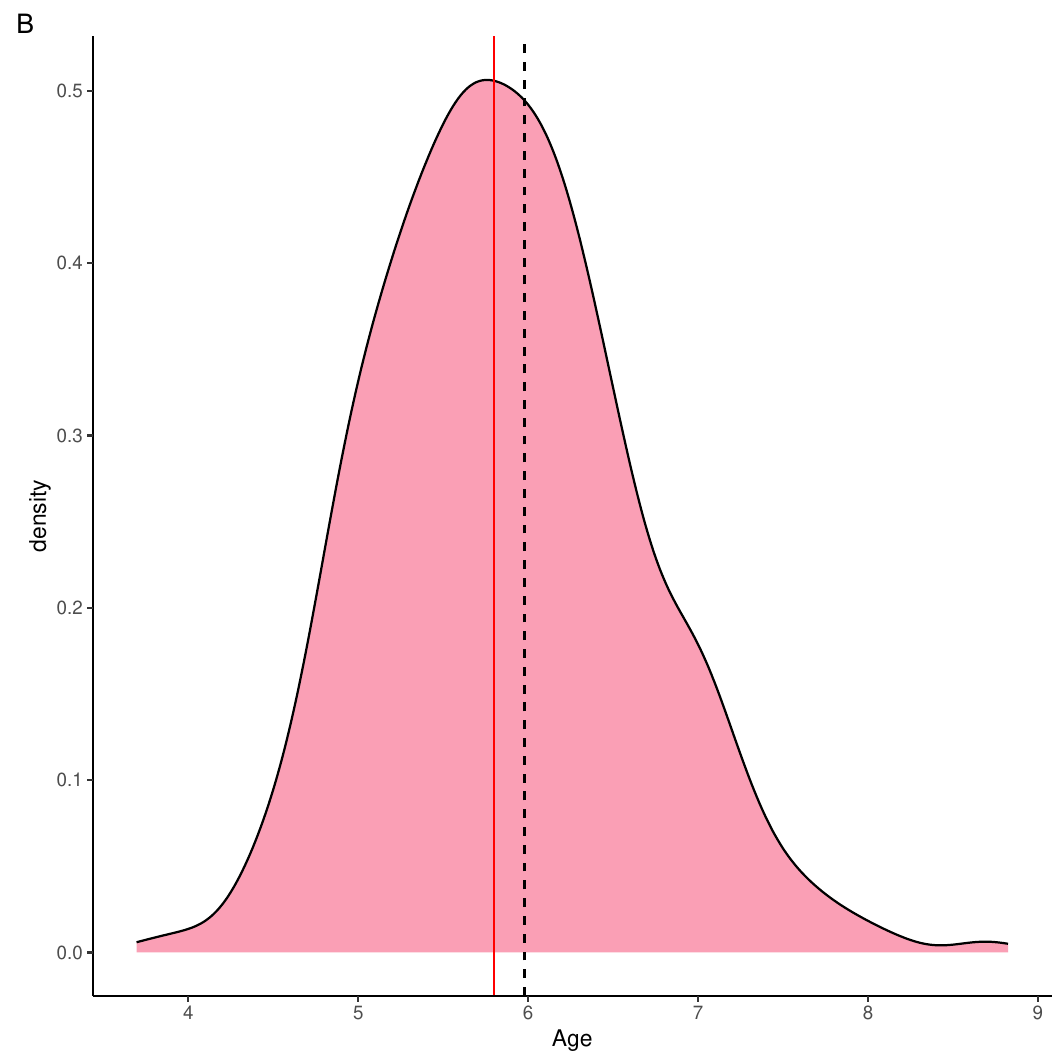}
	\caption{Posterior distribution of the epigenetic age (pink area) estimated from PedBE (\textbf{A}) and Horvath skin and blood (\textbf{B}) clocks, using \emph{CpG predictions} at 6 years for the same illustrative individual. The true age is displayed as a dashed black line, while the red vertical line represents the epigenetic age estimated from the \emph{true CpG values} at 6 years.}
	\label{fig:illu_age_pred}
\end{figure}

\begin{figure}
	\centering
         \includegraphics[width = 0.49\textwidth]{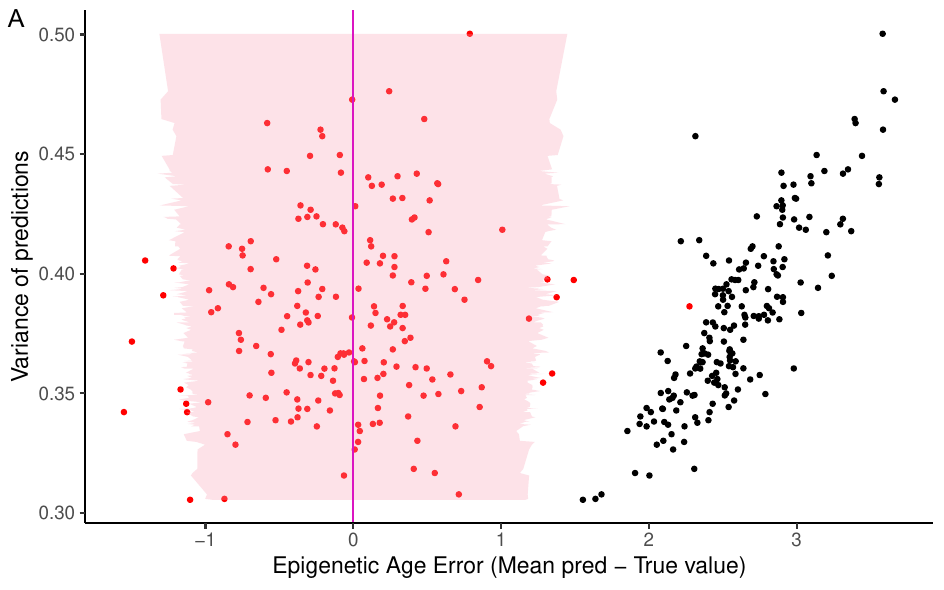}
        \includegraphics[width = 0.49\textwidth]{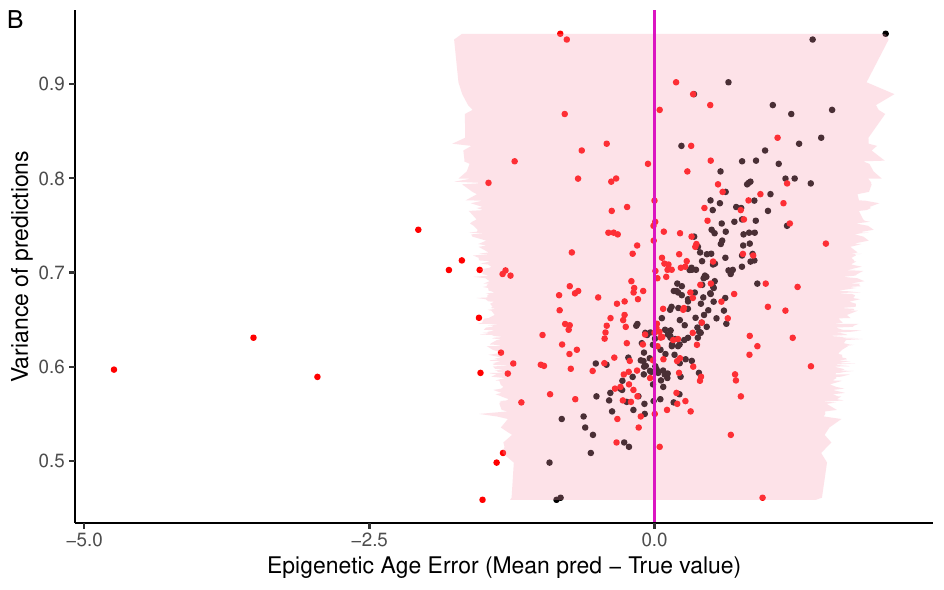}
        \caption{Illustration of the bias between true age and epigenetic age when estimated from the PedBE clock (\textbf{A}) compared with the Horvath skin\&blood clock (\textbf{B}). The predicted mean ages are used as a reference and displayed as a purple line; the pink region corresponds to the associated 95\% credible intervals; each red dot corresponds to the error with the epigenetic age computed using \emph{true observed methylation values}. Each black dot corresponds to the error with \emph{the true age} at the time of data collection.}
        \label{fig:PedBE_bias}
\end{figure}

\begin{figure}
	\centering
        \includegraphics[keepaspectratio, width = \textwidth, height = 0.45\textheight]{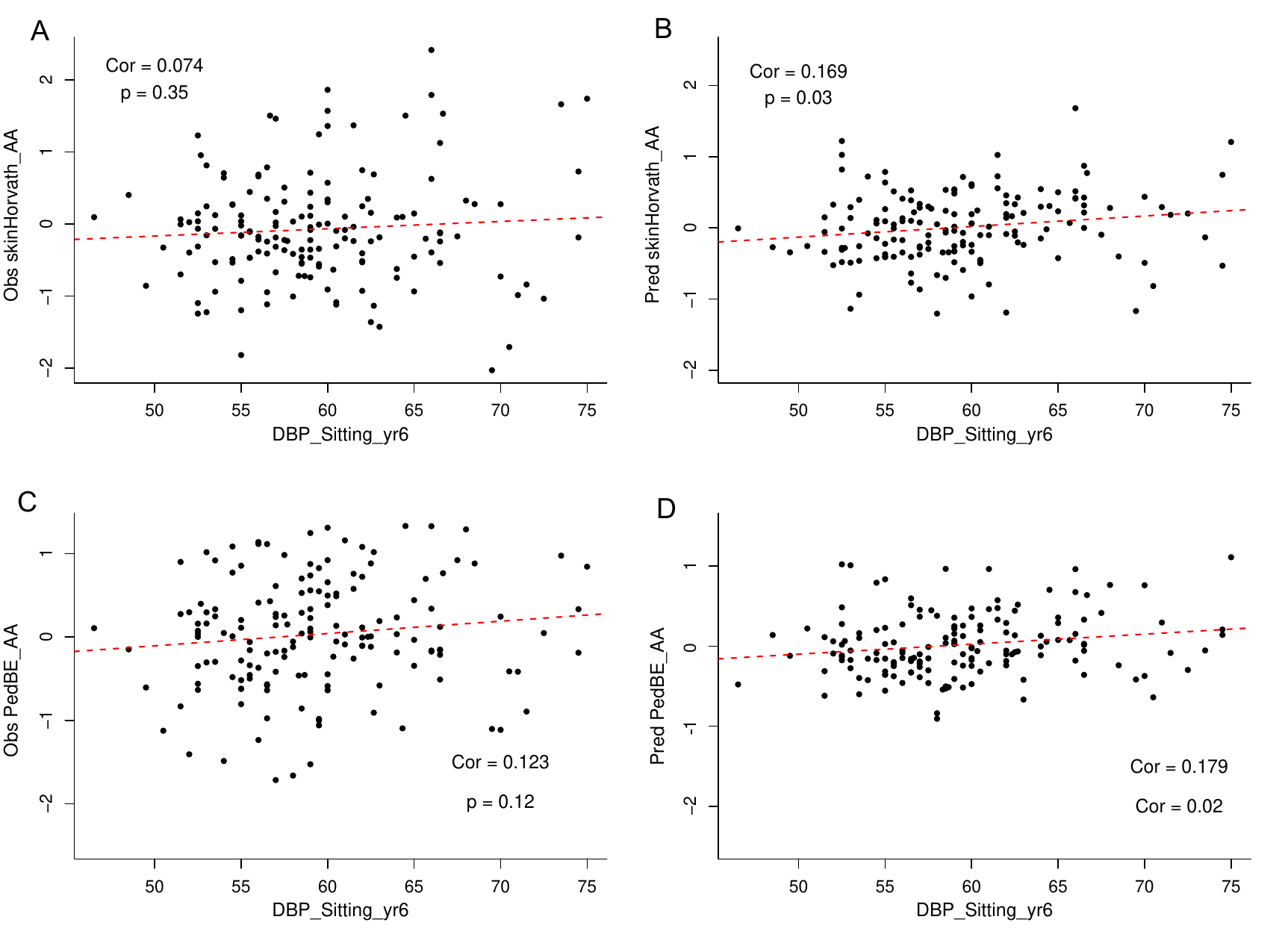}
	\caption{Scatter plot of age acceleration against diastolic blood pressure measured at 6 years old}
	\label{fig:dbp_assoc}
\end{figure}

\begin{figure}[bp!]
\begin{center}
\begin{tikzpicture}
  % Define nodes
  
  \node[obs]                               (y) {$y_i^j$};
  
  \node[latent, below=of y, xshift= 0](mu){$\mu_0$};
  \node[const, below=of mu, xshift=-0.5cm] (m) {$m_0$};
  \node[const, below=of mu, xshift= 0.5cm] (t0) {$\theta_0$};

  \node[latent, left=of y, xshift=0cm] (f) {$f_i$};
  \node[const, above=of f, xshift= 0cm]	   (ti) {$\theta_i$};
  
  \node[latent, right=of y, xshift= 0cm] (g) {$g_j$};
  \node[const, above=of g, xshift= 0cm]	   (tj) {$\theta_j$};
    
  \node[latent, above= 0.5cm of y]           (e) {$\epsilon_i^j$};
  \node[const, above= of e]		   (s) {${\sigma_i^j}^2$};

  \factor[below=of mu] {mu-f} {left:$\mathcal{N}$} {m,t0} {mu} ;
  \factor[above=of f] {f-f} {right:$\mathcal{N}$} {ti} {f} ;
  \factor[above=of g] {g-g} {left:$\mathcal{N}$} {tj} {g} ;
  \factor[above=of e] {e-f} {left:$\mathcal{N}$} {s} {e} ;

  % Connect the nodes
  \edge {f,mu,e, g} {y} ;

  % Plates
  \tikzset{plate caption/.style={caption, node distance=0, inner sep=0pt,
        below left=5pt and 0pt of #1.south,text height=1.2em,text depth=0.3em}}
  \plate {} {(f)(y)(e)(ti)(s)} {$\forall i \in \mathcal{I}$} ;
  \tikzset{plate caption/.style={caption, node distance=0, inner sep=0pt,
        below right=5pt and 0pt of #1.south,text height=1.2em,text depth=0.3em}}
  \plate {} {(g)(y)(e)(tj)(s)} {$\forall j \in \mathcal{P}$} ;
  %\plate {} {(mu)(y)(yx.north west)(yx.south west)} {$M$} ;

\end{tikzpicture}
\caption{Graphical model of dependencies between variables in the multi-mean Gaussian processes model.}
\label{graph_model} 
\end{center}     
\end{figure}
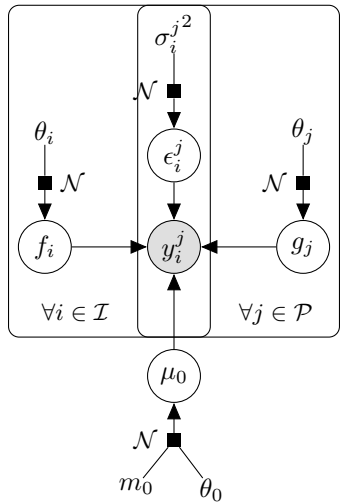

\begin{table}
    \caption{Average (sd) values of RMSE and $CIC_{95}$ for predicted epigenetic age compared with epigenetic age computed from observed methylation value and recorded chronological age at 6 years for 188 testing individuals, using Horvath skin\&blood and PedBE clocks.}
    \centering
\begin{tabular}{c|cc|cc|}
\cline{2-5}
    & \multicolumn{2}{c|}{Epigenetic age} & \multicolumn{2}{c|}{True age} \\
    & RMSE              & $CIC_{95}$      & RMSE           & $CIC_{95}$   \\ \hline
\multicolumn{1}{|c|}{skin\&blood} & 0.70 (2.03)   & 95.20 (21.41) 
    & 0.328 (0.54) &  99.50 (7.29) \\
\multicolumn{1}{|c|}{PedBE}   & 0.36 (0.59)    & 93.6 (24.5)    & 4.09 (1.65) &  4.79 (21.4) \\ \hline
\end{tabular}
    \label{tab:eval_age}
\end{table}

\begin{table}
\caption{Demographic Information of training and testing samples. }
\includegraphics[width = \textwidth]{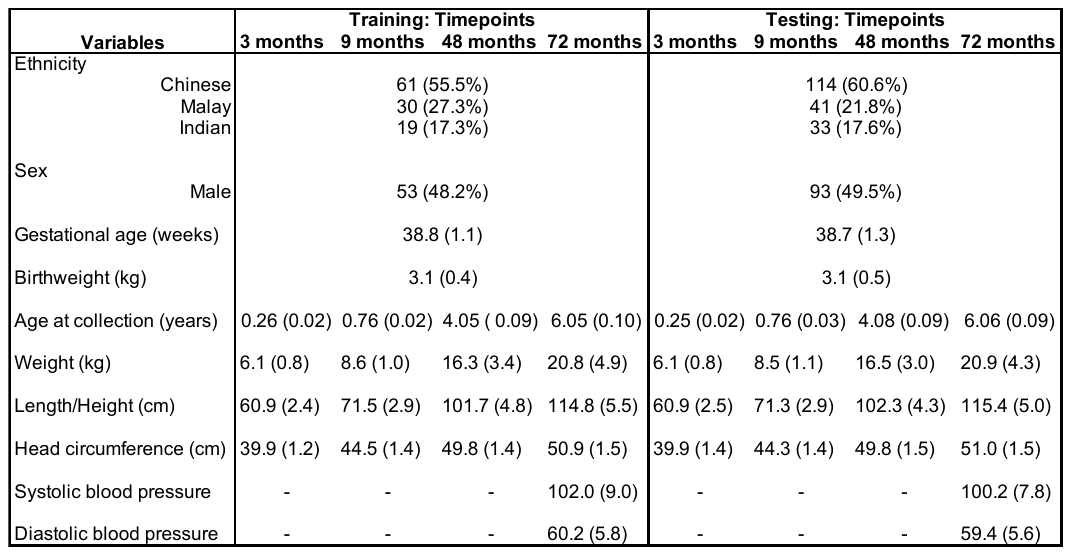}
\label{tab:demo_table}
\end{table}

\end{document}